\documentclass[twocolumn,aps, prd]{revtex4}
\usepackage{amssymb}
\usepackage{graphicx}

\newcommand{\half}{{1\over2}}

\newcommand{\bra}[1]{\langle #1|}
\newcommand{\ket}[1]{|#1\rangle}

\begin{document}

\title{Trans-Planckian Issue  in the Milne Universe}

\author{Pascal M.~Vaudrevange$^{(\rm a, b)}$ and Lev Kofman$^{(\rm a)}$} 
\affiliation{$^{\rm a}${\it Canadian Institute for Theoretical Astrophysics, University of Toronto, 60 St. George St., Toronto, ON M5S 3H8, Canada}\\
  $^{\rm b}${\it Department of Physics, University of Toronto, 60 St. George St., Toronto, ON M5S 1A7, Canada}\\
  {\rm E-mail: \texttt{pascal@physics.utoronto.ca, kofman@cita.utoronto.ca}}
}
\date{\today}

\begin{abstract}
The ``trans-Planckian'' challenge in cosmology appears when we trace
the present physical wavelengths of fluctuations backwards in
time. They become smaller and smaller until crossing the Planck scale
where conventional QFT is challenged, so that unknown ultraviolet
physics may be traced in the observable cosmological
fluctuations. Usually this issue is addressed in the inflationary
context, but trans-Planckian reasoning is much broader. We examine
this logic in a simple example of scalar quantum field theory in the
expanding and contracting Milne universes, where wavelengths of the
eigenmodes are red- or blue-shifted. Trans-Planckian modifications of
QFT should result in a UV-dependent VeV of the energy momentum tensor
of a scalar field in the Milne universe. On the other hand, the Milne
universe is another coordinate systems of flat Minkowski space-time,
and the covariant energy momentum tensor should be the same (but
vacuum-dependent) in different coordinates of flat space time. We
explicitly demonstrate that in conventional QFT the energy momentum
tensor, choosing the adiabatic vacuum, is identical to zero in
Minkowski coordinates, and remains zero in the contracting Milne
universe (due to non-trivial cancellations of contributions from
particles which appear in the accelerating frame and from vacuum
polarization there). In contrast to this, the trans-Planckian
modification of the energy momentum tensor is not motivated. We
provide a similar argument for the expanding Milne universe, where the
energy momentum tensor in the conformal vacuum is non-zero. Similar
arguments are applicable for other cosmological models  where the
curvature is much lower  than Planckian which leads to conflicts with
trans-Planckian considerations.
\end{abstract}

\maketitle

\section{Introduction} \label{sec:introduction}

The quantum theory of cosmological fluctuations generated during the
inflationary stage of the very early universe describes the time
evolution of the scalar eigenmodes $\phi_k(t)e^{i \vec{k x}}$.  The
physical momentum of the eigenmodes $\vec p=\frac{\vec k}{a}$ is
red-shifted in an expanding FRW universe with increasing scalar factor
$a(t)$. If one takes a certain wavelength (of cosmological
fluctuations) today and evolves it backwards in time, due to expansion
this scale will shrink to smaller and smaller values and at one point
will become smaller than the Planckian length. In other words, there
are length scales visible today that have been below Planck length at
some point in the past. This effect is especially dramatic during
inflation, where the scalar factor increases exponentially.
Ultimately, at some instance the physical momentum becomes equal to
the Planckian mass scale $M_p$. After this point, the quantum field
theory approach to fluctuations in an expanding universe (say, during
inflation) should be replaced by a theory incorporating quantum
gravity (say, string theory) which is not yet available. We found the
earliest written traces of the trans-Planckian problem in
\cite{cs89}. The trans-Planckian challenge was articulated in
\cite{trans1}, and discussed in many papers, see e.g \cite{trans}.
\begin{figure}[b]
  \begin{center}
  \includegraphics[scale=.5]{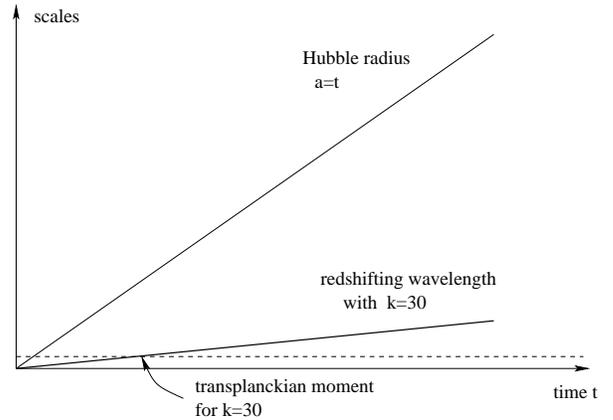}
  \end{center}
  \caption{Redshift of the wavelength $\lambda=\frac{2\pi}{k} a(t)$ in
the Milne universe. The horizontal dashed line represents the Planck
scale $l_p$. A given wavelength for example $k=30$ crosses the
Planckian scale at its trans-Planckian moment $t_k$.}
  \label{fig:1}
\end{figure}

Suppose that trans-Planckian effects alter the result of the
conventional QFT at inflation. Despite an unclear notion about the
microscopic theory of the effect, there is, however, a convenient
phenomenological encoding for it in terms of Bogolyubov coefficients
\cite{Danielsson}.
Indeed, as far as the physical momentum of the mode is below $M_p$,
QFT is applicable. Instead of the vacuum being the positive frequency
eigenmode $f_k(\tau)=\frac{1}{2\omega_k} e^{-i \omega_k \tau}$, one
can use the Bogolyubov coefficients $A_k, B_k$ to describe the mode
function of the initial state
\begin{equation}
\label{mode}
f_k(\tau) \to A_k f_k(\tau)+ B_k f_k^*(\tau) \ , \,\, \vert
A_k \vert^2-\vert B_k \vert^2 \ .
\end{equation}
UV physics, if any, is encoded in the $B_k$.
The trans-Planckian effect looks pretty universal for any expanding FRW universe.
Consider the Milne universe which is a hyperbolic space with FRW type metric
\begin{equation} 
  \label{eqn:metric:milne:real}
  ds^2=dt^2-a(t)^2 \left(dr^2+\sinh^2r\left(d\theta^2+\sin^2\theta d\phi^2\right)\right) \ ,
 \end{equation}
where $a(t)=t$ is the scalar factor. This is the co-moving coordinate
system of the kinematic Milne model, which represents a medium of
probe particles without gravity freely moving from the origin in empty
space-time.

In this model, physical wavelengths are red-shifted, so the reasoning
for a trans-Planckian effect should also be applied here, see
Figure~\ref{fig:1}. However, the Milne universe is in fact another
coordinate system of flat Minkowski space-time, where QFT can be
treated analytically in great details, see
e.g. \cite{Birrell:ix}. Therefore QFT in the Milne universe can be
used as a convenient ground to address the trans-Planckian effect. We
will focus on the VeV of the energy-momentum tensor of a test scalar
field, $\langle T_{\mu}^{\nu}\rangle$, which can be calculated in both
coordinate systems of the flat space-time: in the usual Minkowski
coordinates, and in the Milne coordinates
(\ref{eqn:metric:milne:real}).  For a given choice of vacuum the
answer for the covariant energy-momentum tensor $\langle
T_{\mu}^{\nu}\rangle$ should be the same. However, for the calculation
of $\langle T_{\mu}^{\nu}\rangle$ in the Milne coordinates,
trans-Planckian effects (in terms of $ A_k$ and $B_k$) can be included
and can alter the result. We will consider this as a test to the
trans-Planckian prescription.

Traditionally, the trans-Planckian problem is considered in the
context of an expanding universe, where $ A_k$ and $B_k$ are inherited
from the past. However, we can also consider the trans-Planckian
problem for a contracting universe, where the horizon is still much
bigger than Planckian size while the wavelengths are already
blue-shifted below the Planckian scale. This takes place in the
contracting phase of the Milne universe (similar to the
Figure~\ref{fig:1} but with reverse time direction). For higher
momenta ($k \gg 1$), the ``trans-Planckian'' moment of time where a
given wavelength crosses the Planckian scale occurs at time $t \gg t_p
\sim 10^{-42}$ sec, and where we expect QFT in flat space-time to be
valid.

Therefore we can calculate the VeV of the energy-momentum tensor at
the ``trans-Planckian'' time in Minkowski coordinates which obviously
vanishes
\begin{equation}\label{min}
  \langle T_{\mu}^{\nu}\rangle=0 \ .
\end{equation}
Our goal is to calculate $\langle T_{\mu}^{\nu}\rangle$ in the
contracting Milne universe (\ref{eqn:metric:milne:real}) with and
without trans-Planckian contribution and compare it to the correct
result (\ref{min}). The calculation of $\langle T_{\mu}^{\nu}\rangle$
in the contracting Milne universe is technically easier than that in
the expanding Milne universe (although still quite tedious). Therefore
we first consider the problem in the contracting Milne universe where
methods of calculations will be introduced, and then extend the
results for the expanding Milne universe. We credit \cite{Sahni:vg}
where $\langle T_{\mu}^{\nu}\rangle_{1/2}$ for the spin 1/2 field in
the contracting Milne universe was calculated, and we will extend its
method to the case of the scalar field, and the book
\cite{Grib:1980atom}, where $\langle T_{\mu}^{\nu}\rangle$ for a
scalar field in the contracting Milne universe was calculated by a
different method.

We also consider $\langle T_{\mu}^{\nu}\rangle$ in the expanding Milne
universe.  Since the time $t$ of the Milne coordinates and $\tilde t$
of the Minkowski coordinates are connected by non-linear
transformations, the vacuum choice for the expanding Milne universe is
different from that in the contracting Milne universe.  This is
related to the choice of the conformal vacuum vs. adiabatic vacuum
\cite{Birrell:ix}. As a result, $\langle T_{\mu}^{\nu}\rangle$ in the
Minkowski coordinates for this vacuum is non-zero and corresponds to
an integration over the thermal spectrum of particles seen by an
accelerating observer. On the other hand in an expanding Milne
universe, in addition to this VeV, we may include a potential
trans-Planckian contribution to $A_k, B_k$, which will alter the
expected result.

The plan of this note is as follows. Section 2 contains a short
introduction to the Milne universe. Section 3 is devoted to the
general QFT in the FRW type universes, including the Milne universe.
Section 4 gives a brief outline of the calculation of $\langle
T_{\mu}^{\nu}\rangle$ in contracting Milne universe, while
calculational details are collected in the appendix. Section 5
contains the extension of the results to the expanding Milne universe.
In section 6 we discuss the challenge to the trans-Planckian challenge.

\section{The Milne Universe}
\label{sec:2}

\begin{figure}
  \includegraphics[scale=.4]{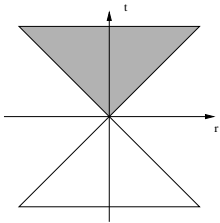}
  \caption{The patches of Minkowski space covered by the Milne metric. Upper (shadow)
patch corresponds to the expanding universe, while the lover (empty) patch corresponds
to the contracting  universe}
  \label{fig:milne}
\end{figure}

The metric of the Milne universe is a FRW type metric
(\ref{eqn:metric:milne:real}) or written in conformal coordinates
\begin{eqnarray}
  \label{eqn:metric:milne:conf}
  &ds^2=a(\eta)^2\left(d\eta^2-dr^2-\sinh^2r\left(d\theta^2+\sin^2\theta d\phi^2\right)\right),&\quad\\
  &a(\eta)=e^\eta,&\nonumber
\end{eqnarray}
where $\eta$ is the conformal time.  For the Milne metric the
curvature tensor vanishes $ R^\mu_{\phantom{\mu}\nu\sigma\rho}=0$, so
it covers a portion of flat space-time. It is related to Minkowski
space-time
\begin{eqnarray}
  ds^2&=&d\tilde{t}^2-d\tilde{r}^2-\tilde{r}^2\left(d\tilde{\theta}^2+\sin^2\tilde{\theta}d\tilde{\phi}^2\right),
\end{eqnarray}
by a coordinate transformation 
\begin{eqnarray}
  \tilde{t}=t\cosh r,&&\tilde{r}=t\sinh r,\\
  \tilde{\theta}=\theta,&&\tilde{\phi}=\phi,\nonumber
\end{eqnarray}
covering the patch $\tilde{t}^2-\tilde{r}^2>0$.  The upper cone $t >
0$ corresponds to an expanding universe, the lower one $t< 0$
corresponds to the contracting universe, see Figure~\ref{fig:milne}.

The conformal properties of the Milne universe are important for the
choice of the vacuum, which is related to global time-like Killing
vectors. Indeed, there is a useful theorem
\cite{Candelas:gf,Birrell:ix}: If for two conformally related
conformally flat space-times $M_1,M_2$ with $M_2$ flat there exists a
diffeomorphism between a global Cauchy hypersurface $\sigma_1$ of
$M_1$ and a global Cauchy hypersurface $\sigma_2$ of $M_2$, then there
exists also a correspondence between the global time-like Killing
vector fields of $M_1$ and $M_2$.

Taking the Milne Universe as $M_1$, it was found \cite{Candelas:gf}
(through mapping both space-times to the Einstein Universe) that these
conditions are fulfilled when taking $M_2$ to be Rindler space with
metric
\begin{eqnarray}
  \label{eqn:metric:rindler}
  &ds^2=e^{2a\xi}\left(d\eta^2-d\xi^2\right)-dy^2-dz^2&,
\end{eqnarray}
which possesses two global time-like Killing vector fields
\begin{eqnarray}
  &\partial_\eta,&\\
  &e^{-a\xi}\cosh(a\eta)\,\,\partial_\eta-e^{-a\xi}\sinh(a\eta)\,\,\partial_\xi,& \nonumber
\end{eqnarray}
corresponding to the conformal vacuum and the adiabatic vacuum. 

So we can deduce that there are also two global time-like Killing
vector fields for the Milne Universe, one of them defining the
conformal vacuum, and the other the adiabatic vacuum. The
corresponding conformal Killing vector fields in the Milne metric
(\ref{eqn:metric:milne:conf}) are
\begin{eqnarray}
  &\Sigma^{(1)}_{\mu}: \,\,\, \partial_\eta,&\\
  & \Sigma^{(2)}_{\mu}:    \,\,\, e^{-\eta}\cosh r \,\,\partial_\eta-e^{-\eta}\sinh r \,\,\partial_r \ ,&\nonumber
\end{eqnarray}
where $\eta, r$ are now the coordinates of the metric (\ref{eqn:metric:milne:conf}). 

We are faced with the choice which vacuum to take. It turns out that
natural choice of vacuum for the contracting Milne universe is the
adiabatic vacuum, associated with the Killing vector $
\Sigma^{(2)}_{\mu}$.
%$e^\eta\cosh r \,\,\partial_\eta+e^\eta\sinh r \,\,\partial_r$.
The adiabatic vacuum also corresponds to the usual vacuum in the
Minkowski coordinates. Indeed, this Killing vector $
\Sigma^{(2)}_{\mu}$ is nothing but $\partial_{\tilde t}$. The
conformal vacuum associated with the other Killing vector $
\Sigma^{(1)}_{\mu}$ is the natural choice of vacuum for the expanding
Milne universe.

\section{Quantum Field Theory in FRW space-times}
\label{sec:3}

Let us recall the basics of the QFT of a free massive real scalar
field in a FRW background. The equation of motion for the scalar field
is given by
\begin{eqnarray}
  \label{eqn:eomphi}
  \left(\Box+m^2-{R\over6}\right)\phi&=&0,
\end{eqnarray}
where $\Box=D_\mu D^\mu$, $m$ is the mass of the scalar field and $R$
is the Ricci scalar. Using conformal time in the FRW universe and
making the Ansatz
\begin{eqnarray}
  \phi(x^\mu)&=&a(\eta)^{-1} \int\!\!d\mu(\lambda, l,m)\,u_\lambda(\eta) \Psi_J(\vec{x}),
\end{eqnarray}
where $\Psi_J$ are eigenfunctions of the spatial Laplacian
$\Delta^{(3)}\Psi_J=\lambda^2\Psi_J$ with quantum numbers
$J=\{\lambda,l,m\}$ and $d\mu(J)$ is the measure over the quantum
numbers, we obtain the following equation for the modes $u_\lambda$
\begin{eqnarray}\label{eigen}
  \ddot{u}_\lambda+\omega^2 u_\lambda=0,&&\omega^2=\lambda^2+m^2a^2,
\end{eqnarray}
where $\dot{u}_\lambda=\partial_\eta u_\lambda$ is the derivative with
respect to conformal time. We quantize the field
\begin{eqnarray}
  \phi(x^\mu)&=&{1\over \sqrt{2}} \int {d\mu(J)\over a(\eta)}\Big(u_\lambda(\eta)\Psi_J(\vec{x})\,a_J\\\nonumber
  &&\phantom{{1\over \sqrt{2}} \int {d\mu(J)\over a(\eta)}abc}+ u_\lambda^*(\eta) \Psi_J^*(\vec{x})\, a_J^\dagger\Big),
\end{eqnarray}
where $a_J^\dagger, a_J$ are the creation and annihilation operators
for particles and antiparticles respectively. Using orthogonality
relations for the eigenmodes of the spatial Laplacian it can be shown
that the $00$-component of the normal ordered energy momentum tensor
for a scalar field is given by
\begin{eqnarray}
  \label{eq:Tno}
  \bra{0}:T_{0}^0:\ket{0}&=&{1\over\pi^2a^4}\int_0^\infty\!\!d\lambda\,\lambda^2\omega s_\lambda,
\end{eqnarray}
where 
\begin{eqnarray}\label{s}
  s_\lambda&=&{1\over2\omega}\Big(|\dot{u}_\lambda|^2+\omega^2|u_\lambda|^2-\omega\Big) \ .
\end{eqnarray}

Now we have to specify the vacuum $\ket{0}$ and the eigenmodes
$u_\lambda$. The general solution to Equation (\ref{eigen}) is
\begin{eqnarray}\label{gen}
  u_\lambda&=&c_1 H_{i\lambda}^{(1)}(\mu)+c_2 H_{i\lambda}^{(2)}(\mu)\ ,
\end{eqnarray}
where we defined $\mu=ma$ with $a=t=e^\eta$. $H_\nu^{(1,2)}(z)$ are
the Hankel functions.

Now consider the contracting Milne universe. Initial conditions shall
be defined at $t \to \infty$.  For $\mu\rightarrow\infty$ we find a
normalized positive energy solution
\begin{eqnarray}
  \label{eqn:hankel}
  u_\lambda&=&{\sqrt{\pi}\over2}e^{{\pi\over2}\lambda} H_{i\lambda}^{(2)}(\mu)\ .
\end{eqnarray}
The correct normalization follows from $\displaystyle (\phi_\lambda,\phi_\nu)=
-\int_\Sigma\!d\Sigma^\mu\sqrt{-g_\Sigma}(\phi_\lambda\partial_\mu\phi_\nu^*-\phi_\nu^*\partial\phi_\mu)\stackrel{!}{=}\delta_{\lambda,\nu}$
which translates into $\displaystyle u_\lambda\partial_t
u_\nu^*-u_\nu^* \partial_t u_\lambda=i$ with $\Sigma^\mu$ being a
time-like conformal Killing vector field orthogonal to the 3-surface
of integration \cite{Mamaev:rus}.

The choice (\ref{eqn:hankel}) of the eigenmode corresponds to the
adiabatic vacuum. Physically, the adiabatic vacuum corresponds to a
vacuum which comes closest to being Minkowski, i.e. it should become
Minkowski in the limit of a very slowly changing geometry. In general
this vacuum is associated with the WKB-type mode solutions of the
equation of motion
\begin{equation}\label{WKB}
  u_\lambda={1\over\sqrt{2 W_\lambda}} e^{-i\int^\eta\!\!d\eta^\prime W_\lambda(\eta^\prime)} \ ,
\end{equation}
where $W$ satisfies the non-linear equation
\begin{equation}\label{WKB1}
  W_\lambda(\eta)^2=\omega^2-{1\over2}\left({\ddot{W}_\lambda\over W_\lambda}-{3\over2}{\dot{W}_\lambda^2\over W_\lambda^2}\right) \ .
\end{equation}
Taking the limit of slowly varying scale factor $a$ or rather
equivalently the limit of large $t$ (or $\eta$), we can approximate
$\omega=me^\eta=W_\lambda$, so that the adiabatic approximation to the
WKB-type solution is given by
\begin{eqnarray}
  u_\lambda&=&{1\over\sqrt{2me^{\eta}}} e^{-ime^\eta}\ ,
\end{eqnarray}
which corresponds to the large $\eta$ limit of the eigenmodes
(\ref{eqn:hankel}). Thus, the adiabatic vacuum $\ket{0_A}$ of the
contracting Milne Universe is the same as the Minkowski vacuum
\cite{Birrell:ix}.

Next, we consider the expanding Milne universe. Initial conditions
shall be defined at $\eta=-\infty$ ($t=0$). We have to select the
solution
\begin{equation} 
  \label{eqn:bessel}
  u_\lambda \to  v_\lambda=\sqrt{\frac{\pi}{2}} \frac{1}{\sqrt{\sinh{\pi\lambda}}} J_{-i\lambda}(\mu) \ ,
\end{equation}
where $J_{i\lambda}(\mu)$ is the Bessel function. Its asymptotic form at
$\eta \to -\infty$ is given by the normalized positive frequency
solution $\frac{1}{\sqrt{2\lambda}}e^{-i\lambda \eta}$. This choice of
eigenmode corresponds to the conformal vacuum.

Indeed, the conformal vacuum is obtained by performing a conformal
transformation of the metric $g_{\mu\nu}$ to the metric
$\tilde{g}_{\mu\nu}$, $g_{\mu\nu}=\Omega^2(x)\tilde{g}_{\mu\nu}$
which changes the equation of motion (\ref{eqn:eomphi}) to $
\tilde{\Box}\tilde{\Phi}=0$, where $\tilde{\Phi}=\Omega^{-1}\Phi$.
The vacuum state associated with the modes $\tilde{u}_\lambda$ of
$\tilde{\Phi}$ corresponds to the conformal vacuum $\ket{0_C}$.

It is instructive to compare the vacua of the contracting (adiabatic)
and expanding (conformal) Milne universe. We can express the orthogonal
set $v_\lambda$ of normalized eigenfunctions of the expanding universe
(\ref{eqn:bessel}) in terms of the set of eigenfunctions $u_\lambda$
of the contracting universe (\ref{eqn:hankel}) introducing Bogolyubov
coefficients
\begin{equation}
  \label{bogo}
  u_{\lambda}(\eta)=\alpha_\lambda v_{\lambda}(\eta)+\beta_\lambda v^*_\lambda(\eta) \ .
\end{equation}
Using the relation between the Bessel and the Hankel functions (see
(\ref{conn})) we find
\begin{equation}
  \label{ab}
  \alpha_\lambda=\frac{e^{{\pi\over2}\lambda}}{\sqrt{2\sinh{\pi\lambda}}} \ , \,\,\, 
  \beta_\lambda=\frac{e^{-{\pi\over2}\lambda}}{\sqrt{2\sinh{\pi\lambda}}} \ .
\end{equation}
In particular, this means that the conformal vacuum in Minkowski
space-time corresponds to excitations of states related to the usual
Minkowski adiabatic vacuum. In other words, in Minkowski coordinates $
\bra{0_A} T_{\mu}^{\nu}\ket{0_A}=0$ but $ \bra{0_C}
T_{\mu}^{\nu}\ket{0_C} \not = 0$.

\section{ $\bra{0_A} T_{\mu}^{\nu(\mathrm{Milne})}\ket{0_A}$ in contracting Milne universe }  
\label{sec:4}

In this section we outline the calculation of the energy momentum
tensor $\bra{0_A} T_{\mu}^{\nu(\mathrm{Milne})}\ket{0_A}$, in the
contracting Milne universe. The starting point is expression
(\ref{eq:Tno}) where we have to substitute the solution
(\ref{eqn:hankel}).

Formally the expression (\ref{eq:Tno}) for the mode functions in the
Milne universe is divergent and needs to be regularized. Normal
ordering in (\ref{eq:Tno}) takes care of the divergence coming from
the zero point energy, but the energy momentum tensor in a curved
space-time features more divergences which are attributed to vacuum
polarization. In FRW type space-times it is most convenient to use the
regularization method of Zel'dovich and Starobinsky
\cite{Zeldovich:1971mw}, which we have adopted. The result is to
replace $s_\lambda$ in (\ref{eq:Tno}) by $s_\lambda-s_2-s_4$ with (see
(\ref{eq:suv}) in the Appendix)
\begin{eqnarray}
  s_2&=&{1\over16}\left({\dot{\omega}\over\omega^2}\right)^2\, \nonumber,\\
  s_4&=&-{3\over256}\left({\dot{\omega}\over\omega^2}\right)^4-
{1\over32}{\dot{\omega}\over\omega^3}{\partial\over\partial\eta}\left[{1\over\omega}{\partial\over\partial\eta}\left({\dot{\omega}\over\omega^2}\right)\right]\nonumber\\
  &&+{1\over64}\left[{1\over\omega}{\partial\over\partial\eta}\left({\dot{\omega}\over\omega^2}\right)\right]^2 \ .
\end{eqnarray}
We illustrate the calculation for the energy density $\rho= T_{00}$,
the other components of $T_{\mu\nu}$ can be either calculated
similarly or from $\rho$, energy conservation $ T_{\mu;\nu}^{\mu}$ and
the vanishing of the conformal anomaly $\bra{0}
T_{\mu}^{\mu}\ket{0}=0$.

For the renormalized energy density we have 
\begin{eqnarray}\label{net}
  T_{0}^{0(\mathrm{ren})}&=&\rho_{\mathrm{vac}}-\rho_0-\rho_1-\rho_2,
\end{eqnarray}
where
\begin{eqnarray}\label{int}
  \rho_{\mathrm{vac}}&=&\lim_{\epsilon\rightarrow0} {1\over2\pi^2a^4}\int_0^\infty\!\!d\lambda\,\lambda^2\Big(|\dot{u}_\lambda|^2+\omega^2|u_\lambda|^2\Big)e^{-\epsilon\lambda}\ ,\nonumber\\
  \rho_0&=&\lim_{\epsilon\rightarrow0}{1\over2\pi^2a^4}\int_0^\infty\!d\lambda\,\lambda^2\omega e^{-\epsilon\lambda} \ ,\nonumber\\
  \rho_{1,2}&=&{1\over\pi^2a^4}\int_0^\infty\!d\lambda\,\lambda^2\omega s_{2,4}\ .
\end{eqnarray}

The first two terms $\rho_{\mathrm{vac}}$ and $\rho_0$ are divergent,
but after regularization they contain finite contributions. The finite
part $\rho_{\mathrm{vac}}$ can be interpreted as a contribution of
particles, seen by the comoving observer in contracting Milne
universe. Terms $\rho_{1,2}$ are finite, and together with the finite
part of $\rho_0$ can be interpreted as the vacuum polarization seen by
the comoving observer.

In order to extract divergences in $\rho_{\mathrm{vac}}$ and
$\rho_0$, we introduce the regularizer $ e^{-\epsilon\lambda}$ with
small dimensionless parameter $\epsilon$.  At the end of our
calculations the final answer will not be dependent on it and we can
send $\epsilon$ to zero. This technical trick is borrowed from
\cite{KSS}.

The results of calculations of the integrals (\ref{int}) detailed in 
the appendix are  the following 
\begin{eqnarray}
  \rho_{\mathrm{vac}}&=&{1\over\pi^2a^4}\Big({3\over\epsilon^4}+{\mu^2\over4\epsilon^2}+{\mu^4\over16}\log{{\epsilon \mu}\over2}\nonumber\\
  &&\phantom{1\over\pi^2a^2}+{\gamma\mu^4\over16}+{\mu^4\over64}+{\mu^2\over48}+{1\over240}\Big)\ ,\\
  \rho_0&=&{1\over\pi^2a^4}\Big({3\over\epsilon^4}+{\mu^2\over4\epsilon^2}+{\mu^4\over16}\log{{\epsilon\mu}\over2}\nonumber\\
  &&\phantom{1\over\pi^2a^2}+{\gamma\mu^4\over16}+{\mu^4\over64}\Big)\ ,\\
  \rho_1&=&{1\over\pi^2a^4}{1\over240},\qquad \rho_2={1\over\pi^2a^4}{\mu^2\over48} \ .
\end{eqnarray}

In the net result (\ref{net}) all divergences are canceled. However,
all finite parts are also canceled so that we end up with
\begin{eqnarray}\label{mmm}
  \bra{0_A}T_{0 (\mathrm{ren})}^{0 (\mathrm{Milne})}\ket{0_A}&=&0.
\end{eqnarray}
All other components of the energy momentum tensor are also zero.

\section{ $\bra{0_C} T_{\mu}^{\nu(\mathrm{Milne})}\ket{0_C}$ in expanding  Milne universe }  
\label{sec:5}

In this section we calculate the energy momentum tensor $\bra{0_C}
 T_{\mu}^{\nu(\mathrm{Milne})}\ket{0_C}$ in the expanding Milne
 universe. Again, we use the formula (\ref{eq:Tno}) where we
 substitute the solution (\ref{eqn:bessel}).

Let us calculate energy density $\rho=\bra{0_C}
T_{00}^{(\mathrm{Milne})}\ket{0_C}$. We have an integral expression
similar to (\ref{eq:Tno}), but with the eigenfunctions $v_\lambda$
instead of $u_\lambda$. In principle, we can apply the method of the
previous section to this case, including regularizing with
$e^{-\epsilon \lambda}$ and extracting divergences. It is easier,
however, to use relationship (\ref{bogo}) between $v_\lambda$ and
$u_\lambda$. Then we obtain
\begin{eqnarray}\label{AC}
  \bra{0_C}T_{0}^{0(\mathrm{Milne})}\ket{0_C}&= &\bra{0_A}T_{0}^{0(\mathrm{Milne})}\ket{0_A}\nonumber\\
  &&+ {1\over\pi^2a^4}\int_0^\infty\!\! d\lambda\,\lambda^2\omega \vert \beta_\lambda \vert^2 (2s_\lambda+1)\nonumber\\
  &&+ \Delta \ ,
\end{eqnarray}
where $s_\lambda$, constructed from $\vert u_\lambda \vert$, is
defined in (\ref{s}). The residual term $\Delta$ is defined in
(\ref{eqn:delta}) and is constructed from $u_{\lambda}^2,
u_{\lambda}^{*2}$.  Before regularization all the divergences are in
the first term $\bra{0_A}T_{00}^{(\mathrm{Milne})}\ket{0_}$.

Expression (\ref{AC}) becomes transparent for large values of $t$ (or
$\eta$), where $s_\lambda \to 1$, $\Delta \to 0$. Regularizing
expression (\ref{AC}) is reduced to regularizing the first term which
we performed in the previous section.  Using (\ref{ab}) for
$\beta_\lambda$, we have the final result
\begin{equation}\label{AC1}
   \bra{0_C}T_{0 (\mathrm{ren})}^{0(\mathrm{Milne})}\ket{0_C}=
{1\over\pi^2a^4}\int_0^\infty\!\! d\lambda\,\lambda^2\omega \frac{1}{e^{2\pi \lambda}-1} \ .
\end{equation}
As expected, in expanding Milne universe energy density is non-zero due to the
choice of the conformal vacuum.

\section{Discussion: Challenge to the Trans-Planckian Challenge}
\label{sec:6}

We will discuss all aspects of the UV physics which may emerge in the 
cosmological models, for instance, the impact of the horizon \cite{Kaloper}
on the trans-Planckian issue.

In an expanding/contracting flat universe, a given wavelength of the
oscillator's eigenmode $e^{i \vec k \vec x}$ is
red-shifted/blue-shifted. The Milne universe has hyperbolic three
dimensional spatial slicing. The spatial eigenmode of this hyperbolic
space is described by the function (\ref{expl}) (see Appendix). For
simplicity we consider the high-frequency modes with $\lambda \gg
1$. In this limit the eigenmode (\ref{expl}) is reduced to simple
standing waves $\sim \cos (\lambda r)$. For these modes we can use the
intuitive red-shifting/blue-shifting picture.

In this section we will discuss how the results (\ref{mmm}),
(\ref{AC1}) will be changed if we apply the trans-Planckian
prescription for the eigenmodes with the wavelengths which were or
will be below the Planckian scale in the contracting/expanding Milne
universe.

In the contracting universe a given wavelength is blue-shifted and will
be shorter than the Planckian scale at its ``trans-Planckian'' moment
$t_k$. If trans-Planckian effects work in this model, then the value
$\bra{0_A}T_{00 (\mathrm{ren})}^{(\mathrm{Milne})}\ket{0_A}$ will
departure from zero. However, the moment $t_k$ is much bigger than the
Planckian time. The moment $t_k$ is not special for observers in the
usual Minkowski coordinate system, where
$\bra{0_A}T_{00(\mathrm{ren}}^{(\mathrm{Mink})}\ket{0_A}$ remains
always zero. We conclude that the trans-Planckian effect does not
emerge in the contracting Milne universe.

Now, let us apply the trans-Planckian prescription to the calculation
of $\bra{0_C}T_{00 (\mathrm{ren})}^{(\mathrm{Milne})}\ket{0_C}$ for
the expanding universe. We simply alter the eigenmode
$u_\lambda(\eta)$ at the moment $\eta_k$, according to (\ref{mode}).
Then the result (\ref{mmm}) will be changed to
\begin{equation}\label{tr}
  \bra{0_C}T_{0(\mathrm{trans})}^{0(\mathrm{Milne})}\ket{0_C}=
      {1\over\pi^2a^4}\int_0^\infty\!\!d\lambda\,\lambda^2\omega  \vert \beta_\lambda+B_\lambda\vert^2 \ .
\end{equation}
However, observers in the usual Minkowski coordinate system should not
see any effects of $B_\lambda$. No trans-Planckian effects emerge in
the expanding Milne universe.

It is clear that our consideration of the trans-Planckian issue in
contracting/expanding cosmologies is much more general than the simple
example of the Milne universe. Consider, for example, the expanding
anisotropic Kasner universe
\begin{equation}\label{kasner}
ds^2=dt^2-t^{2p_1}dx^2-t^{2p_2}dy^2-t^{2p_3}dz^2 \ ,
\end{equation}
with $\sum p_i=\sum p_i^2=1$, so that one of the $p_i$-s is non-positive.
 Suppose space is stretching in two directions
while shrinking in the third direction, $p_3 <0$.  The component of the momentum $k_3$
of the quantum modes $e^{i k_3 z}$ associated with this third direction is blueshifting
(while the universe as a whole is expanding), and at some point passes
the Planckian scale. It is paradoxical to encounter quantum
gravity effects in an expanding Kasner universe!
Notice the special combination of parameters $p_1=p_2=0, p_3=1$. In this case the Kasener metric
(\ref{kasner}) is the product of the two-dimensional Milne universe and  $R^2$ and 
can be transformed to the Minkowski space-time, and we have
another example where the trans-Plankian issue evaporates.
 The resolution  of the paradox: the criterion for the UV (string) physics to become important
is having large values of the curvature terms
$C^{\mu\nu\rho\sigma}_{\mu\nu\rho\sigma}, R^{\mu\nu}_{\mu\nu}, R^2
\sim 1/l_p^4$ but not coordinate effects.

We can compare the situation with another apparent paradox of
inflation, using the ``trans-Planckian'' value of the inflaton filed
$\phi \gg M_p$.  There are no restrictions to use the large $\phi$
values, the physical restrictions are related to the energy density $V \ll
M_p^4$ (where the curvature is sub-Planckian).

\section*{Acknowledgments}

We are grateful to E.~Kolb,  A.~Linde, V.~Mostepanenko, V.~Sahni  and  Yu.~Shtanov for useful
discussions. This research was supported  by the University of
Toronto and NSERC and CIAR.

\section*{Appendix}

In this appendix we describe properties of the eigenfunctions
$u_\lambda \Psi_J(\vec x)$ of (\ref{eqn:eomphi}) and present some
details of the renormalization procedure for the $0-0$ component of
the energy momentum tensor in the conformal and adiabatic vacuum.

\subsubsection*{Properties of Eigenfunctions}
For the hyperbolic case of the open
universe (as in the case of the Milne metric), $J={\lambda,
l,m}$. Ignoring non-normalizable super-horizon modes, we have $0 \le
\lambda \le \infty$, $l=0,1,2, ...$, $m=-l, ..., +l$. The explicit form
of the normalized space-dependent part of the eigenfunction is
\begin{equation}
  \label{expl}
  \Psi_J(\vec x)=\frac{1}{\sqrt{\sinh r  }}\, \frac{\Gamma(i\lambda+l+1)}{\vert\Gamma(i\lambda) \vert}\,
  P^{-l-1/2}_{i\lambda-1/2}(\cosh r) \, Y_{lm}(\theta,\phi) \ ,
\end{equation}
where $P^{\mu}_{\nu}$ are the associated Legendre polynomials, and
$Y_{lm}$ are the spherical harmonics. Let us focus on the high
frequency ($\lambda \gg 1$) asymptotic of $\Psi_J(\vec x)$.  The
asymptotic properties of $P^{-l-1/2}_{i\lambda-1/2}(\cosh r)$ imply
\cite{Bateman}
\begin{equation}
  \label{shift}
  P^{-l-1/2}_{i\lambda-1/2}(\cosh r) \sim \cos(\lambda r) \ .
\end{equation}
The time-dependent part of the eigenfunctions $u_{\lambda}$ is reduced
to the solution of the Bessel equation (\ref{eigen}). 

The two modes $u_\lambda$, corresponding to the adiabatic vacuum, and
$v_\lambda$, corresponding to the conformal vacuum are related
through
\begin{equation}
  \label{conn}
  v_\lambda=\frac{e^{\frac{\pi}{2}\lambda}}{\sqrt{2\sinh{\pi\lambda}}} u_\lambda + 
  \frac{e^{-\frac{\pi}{2}\lambda}}{\sqrt{2\sinh{\pi\lambda}}} u^*_\lambda\ ,
\end{equation}
which can be obtained by using the relation between the Bessel and the
Hankel functions for $\lambda, \mu\in \mathbf{R}$
\begin{equation}
  J_{-i\lambda}(\mu)= \frac{1}{2}
  \left((H_{i\lambda}^{(2)}(\mu))^*+e^{\pi\lambda}H_{i\lambda}^{(2)}(\mu)\right) \ .
\end{equation}
This defines the coefficients $\alpha_\lambda, \beta_\lambda$ in
(\ref{ab}).

\subsubsection*{Renormalizing $\bra{0_A}T_{00}\ket{0_A}$}

To compute the energy momentum tensor, we plug the solutions
(\ref{eqn:hankel}) of the mode equation (\ref{eigen}) into the
definition (\ref{eq:Tno}) of $\bra{0_C}T_{00}\ket{0_C}$. Making use of
the following properties of the Hankel functions \cite{AS}
\begin{eqnarray}
  \left(H_{\nu}^{(1)}(x)\right)^*&=&H_{\nu^*}^{(2)}(x)\ ,\nonumber\\
  H_{-\nu}^{(1)}(z)&=&e^{\nu i\pi}H_{\nu}^{(1)}(z)\ ,\\
  {2\nu\over z}H_\nu^{(i)}(z)&=&H_{\nu-1}^{(i)}(z)+H_{\nu+1}^{(i)}(z) \ ,\nonumber
\end{eqnarray}
we find
\begin{eqnarray}
  \label{eqn:usqr}
  |\dot{u}_\lambda|^2&=&{\pi\over4}\Big[{\mu^2\over2} \Big(H_{i\lambda+1}^{(1)} H_{i\lambda+1}^{(2)} + H_{i\lambda-1}^{(1)} H_{i\lambda-1}^{(2)}\Big)\nonumber\\
  &&\phantom{{\pi\over4}}+\lambda^2H_{i\lambda}^{(1)} H_{i\lambda}^{(2)}\Big]\ ,\\
  |u_\lambda|^2&=&{\pi\over4}H_{i\lambda}^{(1)} H_{i\lambda}^{(2)}\ .\nonumber
\end{eqnarray}
Now we apply 
\begin{eqnarray}
  H_\nu^{(1)}H_\nu^{(2)}&=&{4\over\pi^2}\int_0^\infty\!\!\!\!\!dx\,K_0(2\mu\sinh x) \Big(e^{2\nu x} + e^{-2\nu x}\Big)\ ,\qquad
\end{eqnarray}
where $K_0(z)$ is the MacDonald function, and introduce the
regularizing factor $e^{-\epsilon\lambda}$ to get
\begin{eqnarray}
  \rho_{\mathrm{vac}}&=&{1\over2\pi^2a^4}\int_0^\infty\!\!d\lambda\,\lambda^2\left(|\dot{u}_\lambda|^2+\omega^2|u_\lambda|^2\right)\nonumber\\
  &=&{2\over\pi^3a^4}\int_0^\infty\!dx\,K_0(2\mu\sinh x)\nonumber\\
  &&\phantom{{1\over\pi^3}}\times\Big[\mu^2\cosh(x)^2\int_0^\infty\!d\lambda\,\lambda^2\cos(2\lambda x)e^{-\epsilon\lambda}\nonumber\\
  &&\phantom{{1\over\pi^3a^2}}+\int_0^\infty\!d\lambda\,\lambda^4\cos(2\lambda x)e^{-\epsilon\lambda}\Big]\nonumber\\
  &=&{2\over\pi^3a^4}\int_0^\infty\!dx\,K_0(2\mu\sinh x)\\
  &&\phantom{{2\over\pi^3}}\times\Big[\mu^2\cosh(x)^2 \underbrace{\epsilon{\epsilon^2-12x^2\over(\epsilon^2+4x^2)^3}}_{=:f}\nonumber\\
  &&\phantom{{2\over\pi^3a^4}}+12\underbrace{\epsilon{\epsilon^4-40\epsilon^2x^2+80x^4\over(\epsilon^2+4x^2)^5}}_{=:h}\Big]\ .\nonumber
\end{eqnarray}
Multiplying the fractions $f,h$ by a power of $x^n$ will result in a
zero contribution for $n\geq3$ and $n\geq 5$ respectively. Therefore
we expand $K_0(2\mu\sinh x)\cosh^2x$ to $O(x^3)$ and $K_0(2\mu\sinh
x)$ to order $O(x^5)$ and perform the integrations to obtain
\begin{eqnarray}
  \rho_{\mathrm{vac}}&=&{1\over\pi^2a^4}\Big({3\over\epsilon^4}+{\mu^2\over4\epsilon^2}+{\mu^4\over16}\log{\epsilon\over2}\nonumber\\
  &&\phantom{1\over\pi^2a^4}+{\mu^4\log\mu\over16}+{\gamma\mu^4\over16}+{\mu^4\over64}+{\mu^2\over48}+{1\over240}\Big)\ ,\qquad\quad
\end{eqnarray}
The energy momentum tensor contains divergences even after subtracting
the vacuum energy $\rho_0$ which calculates to
\begin{eqnarray}
  \rho_0&=&{1\over\pi^2a^4}\int_0^\infty\!d\lambda\,\lambda^2{\omega\over2}e^{-\epsilon\lambda}\nonumber\\
  &=&{1\over\pi^2a^4}\\
  &&\times\left({3\over\epsilon^4}+{\mu^2\over4\epsilon^2}+{\mu^4\over16}\log{\epsilon\over2}+{\mu^4\log\mu\over16}+{\gamma\mu^4\over16}+{\mu^4\over64}\right)\ .\nonumber
\end{eqnarray}
To deal with the remaining divergences, we employ the
Zel'dovich-Starobinsky regularization scheme \cite{Zeldovich:1971mw}
which amounts to introducing the variables
$s_\lambda=|\beta_\lambda|^2$ and $u_\lambda-i
v_\lambda=\pm2\alpha_\lambda \beta_\lambda
e^{-2i\int_{\eta_0}^\eta\!d\eta\,\omega}$ obeying
\begin{eqnarray}
  \label{eq:suv}
  \dot{s}_\lambda&=&\half{\dot{\omega}\over\omega} u_\lambda\ ,\nonumber\\
  \dot{v}_\lambda&=&2\omega u_\lambda\ ,\nonumber\\
  \dot{u}_\lambda&=&{\dot{\omega}\over\omega}\left(1\pm 2s_\lambda\right)-2\omega v_\lambda\ .
\end{eqnarray}
We set $\lambda\rightarrow h\lambda, m\rightarrow hm,
\omega\rightarrow h\omega$, expand $s_\lambda, u_\lambda, v_\lambda$
in a series
\begin{eqnarray}
  s_{h\lambda}=\sum_{\ell=1}{1\over h^\ell} s_\ell\ ,\, 
  u_{h\lambda}=\sum_{\ell=1}{1\over h^\ell} u_\ell\ ,\,
  v_{h\lambda}=\sum_{\ell=1}{1\over h^\ell} v_\ell\ ,\quad
\end{eqnarray}
and plug them into (\ref{eq:suv}) to find after some algebra
\begin{eqnarray}
  s_4&=&-\frac{3}{256}\left(\frac{\dot{\omega}}{\omega^2}\right)^4-\frac{1}{32}\frac{\dot{\omega}}{\omega^3}\frac{\partial}{\partial\eta}\left[\frac{1}{\omega}\frac{\partial}{\partial\eta}\left(\frac{\dot{\omega}}{\omega^2}\right)\right]\nonumber\\
  &&+\frac{1}{64}\left[\frac{1}{\omega}\frac{\partial}{\partial\eta}\left(\frac{\dot{\omega}}{\omega^2}\right)\right]^2\ ,\quad s_2=\frac{1}{16}\left(\frac{\dot{\omega}}{\omega^2}\right)^2\ ,\quad
\end{eqnarray}
with all other $s_\ell=0$. Now we can easily calculate
\begin{eqnarray}
  \rho_1&=&{1\over\pi^2 a^4}\int_0^\infty\!d\lambda\,\lambda^2 \omega s_2={1\over\pi^2a^2}{\mu^2\over48}\ ,\nonumber\\
  \rho_2&=&{1\over\pi^2 a^4}\int_0^\infty\!d\lambda\,\lambda^2 \omega s_4={1\over\pi^2a^2}{1\over240}\ .
\end{eqnarray}
Adding up all terms gives
\begin{eqnarray}
  \bra{0_A}T_{0}^{0(\mathrm{ren})}\ket{0_A}&=&\rho_{\mathrm{vac}}-\rho_0-\rho_1-\rho_2=0.
\end{eqnarray}

\subsubsection*{Renormalizing $\bra{0_C}T_{00}\ket{0_C}$}

In order to compute $\bra{0_C}T_{00}\ket{0_C}$ we plug the conformal
eigenmodes (\ref{eqn:bessel}) into (\ref{s}), express the modes
$v_\lambda, v^*_\lambda$ in terms of the modes $u_\lambda,
u^*_\lambda$ (\ref{eqn:hankel}) and find after some algebra
\begin{eqnarray}
  \bar{s}_\lambda&=&\frac{1}{2\omega}\left(|\dot{v}_\lambda|^2+\omega^2|v_\lambda|^2-\omega\right)\nonumber\\
  &=&s_\lambda+ \frac{e^{-\pi\lambda}}{\sinh{\pi\lambda}} \left(s_\lambda + \frac{1}{2}\right)+\tilde{\Delta}\ ,
\end{eqnarray}
where we can identify the second term as $|\beta_\lambda|^2 (
2s_\lambda+1)$ and
\begin{equation}
  \tilde{\Delta}=\frac{1}{4\omega\sinh{\pi\lambda}}\left(\dot{u}_\lambda^2 + \omega^2 u_\lambda^2 + \dot{u}_\lambda^{*2} + \omega^2 u_\lambda^{*2}\right)\ .
\end{equation}
This gives rise to expression (\ref{AC}) for
$\bra{0_C}T_{0}^0\ket{0_C}$ with
\begin{equation}
  \label{eqn:delta}
  \Delta=\frac{1}{\pi^2 a^4}\int d\lambda\, \lambda^2 \omega \tilde{\Delta}\ .
\end{equation}
At late times, the integrand $\tilde{\Delta}$ is rapidly oscillating
so that $\lim_{t\rightarrow\infty}\Delta = 0$ as can be seen by using
the large argument asymptotics of the Hankel functions
\begin{eqnarray}
  \label{eqn:lateHankel}
  \lim_{\mu\rightarrow\infty}H^{(1)}(\mu) &=& \sqrt{\frac{2}{\pi\mu}}e^{i\left(\mu-\frac{1}{2}\nu\pi-\frac{\pi}{4}\right)}\ ,\nonumber\\
  \lim_{\mu\rightarrow\infty}H^{(2)}(\mu) &=& \sqrt{\frac{2}{\pi\mu}}e^{-i\left(\mu-\frac{1}{2}\nu\pi-\frac{\pi}{4}\right)}\ .
\end{eqnarray}
The late time behaviour of $(2s_\lambda+1)\rightarrow1$ can be easily
seen from (\ref{eqn:usqr}) using the asymptotic behaviour of the
Hankel functions (\ref{eqn:lateHankel}).

%%% Bibliography

\end{document}